\documentclass[12pt,APS,preprint]{revtex4}
\usepackage[cmex10]{amsmath}
\usepackage{array}
\usepackage{graphicx}

\begin{document}

\title{Third-order triangular finite elements for waveguiding problems}

\author{E.~Cojocaru} 
\affiliation{Department of Theoretical Physics, Horia Hulubei National Institute of Physics and Nuclear Engineering, Magurele-Bucharest P.O.Box MG-6, 077125 Romania}
\email{ecojocaru@theory.nipne.ro} 
 
\begin{abstract}
Explicit relations of matrices for two-dimensional finite element method with third-order triangular elements are given. They are more simple than relations presented in other works  and could be easily implemented in new algorithms for both isotropic and anisotropic materials. Numerical examples are given comparatively using second-order and third-order triangular elements for problems of wave propagation in rectangular waveguides which have analytic solutions.
\end{abstract}

\maketitle
 
\section{Introduction}
\begin{figure}[!h]
\begin{center}
\includegraphics[width=2.5in]{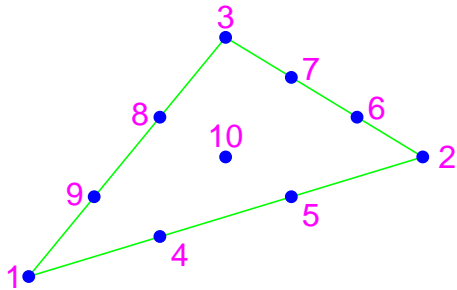}
\caption{Third-order triangular element}
\label{fig:1}
\end{center}
\end{figure}

The finite element method is a widely applicable numerical technique for obtaining approximate solutions to boundary-value problems of mathematical physics \cite{1,2}. Any complex shape of the problem domain can be handed with ease by division into many subdomains, each subdomain being called a finite element. For two-dimensional problems one resorts usually to triangular elements: the first-order triangular element, which requires three nodes, and the second-order triangular element, which requires six nodes. In order to achieve higher accuracy in the finite element solution, two approaches are commonly taken: one resorts to finer subdivision or smaller elements, and the other resorts to higher-order interpolation functions or higher-order elements. Here we are interested in the two-dimensional finite element method with third-order triangular elements, each element requiring ten nodes which are numbered counterclockwise as shown in Fig.\ref{fig:1}. These high-order elements can be successfully employed for the characterization of wave propagation in shielded microstrip transmission lines or integrated circuits with slot lines, when the lines are infinitesimally thin and one needs to place a set of nodes above the line as well bellow the line, as if the line had a finite thickness \cite{2}. Generally, the elemental matrices for high-order elements are determined numerically, but a higher accuracy is assured with explicit expressions. Relations for triangular elements, including the third-order ones, have been presented in \cite{3}. In this paper the elemental matrices for third-order triangles are given in simple, explicit forms which can be easily implemented in different algorithms of waveguiding problems related to both isotropic and anisotropic materials.

\section{General relations}

Let us consider a two-dimensional boundary-value problem defined by the second-order differential equation
\begin{equation}
\label{eq:1}
-\frac{\partial}{\partial x}\Big(\alpha_x\frac{\partial \phi}{\partial x}\Big)- 
\frac{\partial}{\partial y}\Big(\alpha_y\frac{\partial \phi}{\partial y}\Big)+\beta\phi=f,
\quad (x,y)\in\Omega
\end{equation}
where $\phi$ is the unknown function; $\alpha_x,\alpha_y,~\text{and}~\beta$ are known parameters; and $f$ is the source or excitation function. The ordinary two-dimensional Laplace equation, Poisson equation, and Helmholtz equation are special forms of (\ref{eq:1}). For simplicity we consider $f\!=\!0$ and the homogeneous Neumann boundary condition on the boundary enclosing the area $\Omega$. Within each element, $\phi$ can be approximated as
\begin{equation}
\label{eq:2}
\phi^e(x,y)=\sum_{j=1}^{10}N_j^e(x,y)\phi_j^e,
\end{equation}
where $\phi_j^e$ are constant expansion coefficients and $N_j^e(x,y)$ are the interpolation or expansion functions given by \cite{2}
\begin{eqnarray}
\label{eq:3}
N_1^e(x,y)&{}={}&\frac{1}{2}L_1^e(3L_1^e-1)(3L_1^e-2), \nonumber \\
N_2^e(x,y)&{}={}&\frac{1}{2}L_2^e(3L_2^e-1)(3L_2^e-2), \nonumber \\
N_3^e(x,y)&{}={}&\frac{1}{2}L_3^e(3L_3^e-1)(3L_3^e-2), \nonumber \\
N_4^e(x,y)&{}={}&\frac{9}{2}L_1^eL_2^e(3L_1^e-1), \nonumber \\ 
N_5^e(x,y)&{}={}&\frac{9}{2}L_1^eL_2^e(3L_2^e-1),\nonumber \\
N_6^e(x,y)&{}={}&\frac{9}{2}L_2^eL_3^e(3L_2^e-1), \nonumber \\ 
N_7^e(x,y)&{}={}&\frac{9}{2}L_2^eL_3^e(3L_3^e-1), \nonumber \\ 
N_8^e(x,y)&{}={}&\frac{9}{2}L_1^eL_3^e(3L_3^e-1), \nonumber \\
N_9^e(x,y)&{}={}&\frac{9}{2}L_1^eL_3^e(3L_1^e-1), \nonumber \\
N_{10}^e(x,y)&{}={}&27L_1^eL_2^eL_3^e. 
\end{eqnarray}
In the above, the area coordinates $L_j^e$ are given by
\begin{equation}
\label{eq:4}
L_j^e=\frac{1}{2\Delta^e}(a_j+b_jx+c_jy), \qquad j=1,2,3,
\end{equation}
in which $a_j,b_j,~\text{and}~c_j$ are
\begin{eqnarray*}  
a_1&{}={}&x_2y_3-y_2x_3, \qquad b_1=y_2-y_3, \qquad c_1=x_3-x_2, \\
a_2&{}={}&x_3y_1-y_3x_1, \qquad b_2=y_3-y_1, \qquad c_2=x_1-x_3, \\
a_3&{}={}&x_1y_2-y_1x_2, \qquad b_3=y_1-y_2, \qquad c_3=x_2-x_1,
\end{eqnarray*} 
where $x_j$ and $y_j~(j\!=\!1,2,3)$ denote the coordinate values of the vertices and $\Delta^e$ is the area of the $e$th element,\[\Delta^e=\frac{1}{2}(b_1c_2-b_2c_1).\]
With the expansion of $\phi$ given in (\ref{eq:2}), we can proceed to formulate the elemental equations using either the Ritz or Galerkin method \cite{2,4}. As for example, in the Ritz method we formulate the problem in terms of a functional $F(\phi)$ whose minimum corresponds to the differential equation of the boundary-value problem. The functional can be written as
\begin{equation}
\label{eq:5}
F(\phi)=\sum_{e=1}^{M}F^e(\phi^e),
\end{equation}
where $M$ denotes the total number of elements and $F^e$ is the subfunctional corresponding to the $e$th element. For differential equation (\ref{eq:1}), $F^e$ is defined as
\begin{equation}
\label{eq:6}
F^e(\phi^e)=\frac{1}{2}\iint_{\Omega^e}\Big[\alpha_x\Big(\frac{\partial\phi^e}{\partial x}\Big)^2+ 
\alpha_y\Big(\frac{\partial\phi^e}{\partial y}\Big)^2+\beta(\phi^e)^2\Big]\mathrm{d}\Omega,
\end{equation}
where $\Omega^e$ is the domain of the $e$th element. Introducing expression (\ref{eq:2}) for $\phi^e$ and differentiating with respect to $\phi_i^e$ yields
\begin{equation}
\label{eq:7}
\frac{\partial F^e}{\partial\phi_i^e} = \sum_{j=1}^{10}\phi_j^e\iint_{\Omega^e}(\alpha_x
\frac{\partial N_i^e}{\partial x}\frac{\partial N_j^e}{\partial x}+\alpha_y\frac{\partial N_i^e}
{\partial y}\frac{\partial N_j^e}{\partial y} 
+ \beta N_i^eN_j^e)\mathrm{d}\Omega, \qquad i{}={}1,2,\dots 10.
\end{equation}
In matrix form, this can be written as
\begin{equation}
\label{eq:8}
\left\{\frac{\partial F^e}{\partial\phi^e}\right\} = [K^e]\left\{\phi^e\right\},
\end{equation}
where
\begin{equation*}
\left\{\frac{\partial F^e}{\partial\phi^e}\right\} = \left[\frac{\partial F^e}{\partial\phi_1^e}
\frac{\partial F^e}{\partial\phi_2^e}\cdots\frac{\partial F^e}{\partial\phi_{10}^e}\right]^T , \qquad 
\left\{\phi^e\right\} = \left[\phi_1^e\phi_2^e \dots \phi_{10}^e\right]^T,
\end{equation*}
with $T$ denoting a transpose and the elements of the matrix $[K^e]$ given by
\begin{equation}
\label{eq:9}
K_{ij}^e = \iint_{\Omega^e}(\alpha_x\frac{\partial N_i^e}{\partial x}\frac{\partial N_j^e}
{\partial x} + \alpha_y\frac{\partial N_i^e}{\partial y}\frac{\partial N_j^e}{\partial y} 
+\beta N_i^eN_j^e)\mathrm{d}x\mathrm{d}y, \qquad
i,j = 1,2,\dots 10.
\end{equation}

\section{Elemental matrices}

Assuming $\alpha_x,~\alpha_y,~\text{and}~\beta$ are constant in each element, we split $[K^e]$ into three parts
\begin{equation}
\label{eq:10}
[K^e]=\alpha_x[A_x^e]+\alpha_y[A_y^e]+\beta[B^e],
\end{equation}
where the elements of matrices $[A_x^e],[A_y^e],~\text{and}~[B^e]$ are given by
\begin{eqnarray}
\label{eq:11}
A_{xij}^e&{}={}& \iint_{\Omega^e}\frac{\partial N_i^e}{\partial x}\frac{\partial N_j^e}
{\partial x}\mathrm{d}x\mathrm{d}y,  \qquad
A_{yij}^e = \iint_{\Omega^e}\frac{\partial N_i^e}{\partial y}\frac{\partial N_j^e}
{\partial y}\mathrm{d}x\mathrm{d}y,   \nonumber \\
B_{ij}^e&{}={}&\iint_{\Omega^e}N_i^eN_j^e\mathrm{d}x\mathrm{d}y,\qquad i,j{}={}1,2,\dots 10. 
\end{eqnarray}
The integral calculus in (\ref{eq:11}) can be performed analytically by using the following convenient integration formula for the area coordinates \cite{5}
\begin{equation}
\label{eq:12}
\iint_{\Omega^e}(L_1^e)^i(L_2^e)^j(L_3^e)^k\mathrm{d}x\mathrm{d}y = \frac{i!j!k!}{(i+j+k+2)!}2\Delta^e ,
\qquad i,j,k = 0,1,2,3,\dots
\end{equation}
Explicit expressions for the matrix elements $B_{ij}^e$ and $A_{xij}^e$ are given in Tables \ref{tab:1} and \ref{tab:2}, respectively. Both $[B^e]$ and $[A_x^e]$ are symmetric matrices, i.e., $B_{ji}^e\!=\!B_{ij}^e$ and
$A_{xji}^e\!=\!A_{xij}^e$. In Table \ref{tab:2} we used notation $b_{ij}\!=\!b_ib_j$, with $i,j\!=\!1,2,3$. The matrix elements $A_{yij}^e$ are obtained by changing $b_i$ to $c_i~(i\!=\!1,2,3)$ in Table \ref{tab:2}.

\begin{table}[!h]
\begin{center}
\caption{\label{tab:1}Matrix elements $B_{ij}^e~(i,j=1,2,\dots 10)$ are tabulated values multiplied by $(3\Delta^e/2240)$}
\begin{tabular}{|l||r|r|r|r|r|r|r|r|r|r|}
\hline
\hline
$i\backslash j$&\textit{1}
&\textit{2}&\textit{3}&\textit{4}&\textit{5}&\textit{6}&\textit{7}&\textit{8}&\textit{9}&\textit{10}\\
\hline
\hline
\textit{1}&$\frac{76}{9}$&$\frac{11}{9}$&$\frac{11}{9}$&2&0&3&3&0&2&4\\
\hline
\textit{2}&$\frac{11}{9}$&$\frac{76}{9}$&$\frac{11}{9}$&0&2&2&0&3&3&4\\
\hline
\textit{3}&$\frac{11}{9}$&$\frac{11}{9}$&$\frac{76}{9}$&3&3&0&2&2&0&4\\
\hline
\textit{4}&2&0&3&60&-21&-15&-6&-15&30&18\\
\hline
\textit{5}&0&2&3&-21&60&30&-15&-6&-15&18\\
\hline
\textit{6}&3&2&0&-15&30&60&-21&-15&-6&18\\
\hline
\textit{7}&3&0&2&-6&-15&-21&60&30&-15&18\\
\hline
\textit{8}&0&3&2&-15&-6&-15&30&60&-21&18\\
\hline
\textit{9}&2&3&0&30&-15&-6&-15&-21&60&18\\
\hline
\textit{10}&4&4&4&18&18&18&18&18&18&216\\
\hline
\hline
\end{tabular}
\end{center}
\end{table}

\begin{table}[!ht]
\begin{center}
\caption{\label{tab:2}Matrix elements $A_{xij}^e~(i,j=1,2,\dots 10)$ are tabulated expressions multiplied by $[81/(8\Delta^e)]$}
\begin{tabular}{|l||c|c|c|c|c|c|c|c|c|c|}
\hline
\hline
$i\backslash j$&\textit{1} &\textit{2}&\textit{3}&\textit{4}&\textit{5}&\textit{6}&\textit{7}&\textit{8}&\textit{9}&\textit{10}\\
\hline
\hline
\textit{1}&\mbox{$\frac{17b_{11}}{810}$}&\mbox{$\frac{7b_{12}}{1620}$}&\mbox{$\frac{7b_{13}}{1620}$}&\mbox{$\frac{18b_{12}-b_{13}}{540}$}&\mbox{$
\frac{-b_{13}-9b_{12}}{540}$}&\mbox{$\frac{-b_{11}}{540}$}&\mbox{$\frac{-b_{11}}{540}$}&\mbox{$\frac{-b_{12}-9b_{13}}{540}$}&\mbox{$\frac{18b_{13}-b_{12}}{540}$}&\mbox{0}\\
\hline
\textit{2}&&\mbox{$\frac{17b_{22}}{810}$}&\mbox{$\frac{7b_{23}}{1620}$}&\mbox{$\frac{-b_{23}-9b_{12}}{540}$}&\mbox{$\frac{18b_{12}-b_{23}}{540}$}
&\mbox{$\frac{18b_{23}-b_{12}}{540}$}&\mbox{$\frac{-b_{12}-9b_{23}}{540}$}&\mbox{$\frac{-b_{22}}{540}$}&\mbox{$\frac{-b_{22}}{540}$}&\mbox{0}\\
\hline
\textit{3}&&&\mbox{$\frac{17b_{33}}{810}$}&\mbox{$\frac{-b_{33}}{540}$}&\mbox{$\frac{-b_{33}}{540}$}&\mbox{$\frac{-b_{13}-9b_{23}}{540}$}&
\mbox{$\frac{18b_{23}-b_{13}}{540}$}&\mbox{$\frac{18b_{13}-b_{23}}{540}$}&
\mbox{$\frac{-b_{23}-9b_{13}}{540}$}&\mbox{0}\\
\hline
\textit{4}&&&&\mbox{$\frac{b_{33}-b_{12}}{12}$}&\mbox{$\frac{-(b_1-b_2)^2}{60}$}&\mbox{$\frac{-b_{13}}{60}$}&\mbox{$\frac{-b_{13}}{60}$}&
\mbox{$\frac{-b_{23}}{60}$}&\mbox{$\frac{b_{23}}{12}$}&\mbox{$\frac{b_{13}}{10}$}\\
\hline
\textit{5}&&&&&\mbox{$\frac{b_{33}-b_{12}}{12}$}&\mbox{$\frac{b_{13}}{12}$}&\mbox{$\frac{-b_{13}}{60}$}&\mbox{$\frac{-b_{23}}{60}$}&\mbox{$
\frac{-b_{23}}{60}$}&\mbox{$\frac{b_{23}}{10}$}\\
\hline
\textit{6}&&&&&&\mbox{$\frac{b_{11}-b_{23}}{12}$}&\mbox{$\frac{-(b_2-b_3)^2}{60}$}&\mbox{$\frac{-b_{12}}{60}$}&\mbox{$\frac{-b_{12}}{60}$}&\mbox{$\frac{b_{12}}{10}$}\\
\hline
\textit{7}&&&&&&&\mbox{$\frac{b_{11}-b_{23}}{12}$}&\mbox{$\frac{b_{12}}{12}$}&\mbox{$\frac{-b_{12}}{60}$}&\mbox{$\frac{b_{13}}{10}$}\\
\hline
\textit{8}&&&&&&&&\mbox{$\frac{b_{22}-b_{13}}{12}$}&\mbox{$\frac{-(b_1-b_3)^2}{60}$}&\mbox{$\frac{b_{23}}{10}$}\\
\hline
\textit{9}&&&&&&&&&\mbox{$\frac{b_{22}-b_{13}}{12}$}&\mbox{$\frac{b_{12}}{10}$}\\
\hline
\textit{10}&&&&&&&&&&\mbox{$\frac{b_{11}-b_{23}}{5}$}\\
\hline
\hline
\end{tabular}
\end{center}
\end{table} 
In the vector formulation of different waveguiding problems, the subfunctional $F^e$ of the $e$th element in (\ref{eq:6}) is more complicated and the matrix $[K^e]$ in (\ref{eq:10}) contains more elemental matrices. As for example, within a closed waveguide, the magnetic field satisfies the vector differential equation
\begin{equation}
\label{eq:13}
\nabla \times (\frac{1}{\epsilon_r}\nabla \times \mathbf{H})-k_0^2\mu_r\mathbf{H}=0 \quad \textrm{in} \quad \Omega
\end{equation}
and the boundary condition $\hat{n} \times (\nabla \times \mathbf{H})=0$ on $\Gamma_1$, where $\Omega$ denotes the cross section of the structure comprised by the electric wall $\Gamma_1$, $k_0$ is the wave number in vacuum, $\epsilon_r$ and $\mu_r$ are the permittivity and permeability of the structure. With the $z$-dependence of $\mathbf{H}$ as $\mathbf{H}(x,y,z)=\mathbf{H}(x,y)\mathrm{e}^{j(\omega t-k_zz)}$, where $\omega$ is the circular frequency, the functional of (\ref{eq:13}) can be written as \cite{2}
\begin{eqnarray}
\label{eq:14}
F(\mathbf{H})&=&\frac{1}{2}\iint_\Omega [\frac{1}{\epsilon_r}\left(\left|\frac{\partial H_z}{\partial y}+jk_zH_y\right|^2+\left|jk_zH_x+\frac{\partial H_z}{\partial x}\right|^2+\left|\frac{\partial H_y}{\partial x} - \frac{\partial H_x}{\partial y}\right|^2\right) \nonumber \\ &-&k_0^2\mu_r(|H_x|^2+|H_y|^2+|H_z|^2) ]\mathrm{d}\Omega .
\end{eqnarray}
To render this as a real system, we introduce the transformation $h_z=-jH_z$, and with this, (\ref{eq:14}) becomes
\begin{eqnarray}
\label{eq:15}
F(\mathbf{H})&=&\frac{1}{2}\iint_\Omega [\frac{1}{\epsilon_r}\left(\left|\frac{\partial h_z}{\partial y}+k_zH_y\right|^2+\left|k_zH_x+\frac{\partial h_z}{\partial x}\right|^2+\left|\frac{\partial H_y}{\partial x} - \frac{\partial H_x}{\partial y}\right|^2\right) \nonumber \\ &-&k_0^2\mu_r(|H_x|^2+|H_y|^2+|h_z|^2)]\mathrm{d}\Omega
\end{eqnarray} 
The functional can readily be discretized in a standard manner, and the result is
\begin{equation}
\label{eq:16}
\left[\begin{array}{ccc}
A_{xx} & A_{xy} & A_{xz}\\
A_{yx} & A_{yy} & A_{yz}\\
A_{zx} & A_{zy} & A_{zz}
\end{array} \right]
\left\{\begin{array}{l}H_x\\H_y\\h_z\end{array}\right\} = k_0^2 \left[
\begin{array}{ccc}
B_x & 0 & 0 \\
0 & B_y & 0 \\
0 & 0 & B_z
\end{array} \right]
\left\{\begin{array}{l}H_x\\H_y\\h_z\end{array}\right\},
\end{equation}
where the matrices are assembled from their corresponding elemental matrices, given by
\begin{equation*}
[A_{xx}^e]= \iint_{\Omega^e}\frac{1}{\epsilon_r}\left(\frac{\partial \{N^e\}}{\partial y}\frac{\partial \{N^e\}^T}{\partial y}+k_z^2\{N^e\}\{N^e\}^T\right)\mathrm{d}\Omega
\end{equation*}

\begin{equation*}
[A_{yy}^e]=\iint_{\Omega^e}\frac{1}{\epsilon_r}\left(\frac{\partial \{N^e\}}{\partial x}\frac{\partial \{N^e\}^T}{\partial x}+k_z^2\{N^e\}\{N^e\}^T\right)\mathrm{d}\Omega
\end{equation*}

\begin{equation*}
[A_{zz}^e]=\iint_{\Omega^e}\frac{1}{\epsilon_r}\left(\frac{\partial \{N^e\}}{\partial x}\frac{\partial \{N^e\}^T}{\partial x}+\frac{\partial \{N^e\}}{\partial y}\frac{\partial \{N^e\}^T}{\partial y}\right)\mathrm{d}\Omega
\end{equation*}

\begin{equation*}
[A_{xy}^e]=[A_{yx}^e]^T=-\iint_{\Omega^e}\frac{1}{\epsilon_r}\frac{\partial \{N^e\}}{\partial y}\frac{\partial \{N^e\}^T}{\partial x}\mathrm{d}\Omega
\end{equation*}

\begin{equation*}
[A_{yz}^e]=[A_{zy}^e]^T=\iint_{\Omega^e}\frac{k_z}{\epsilon_r}\{N^e\}\frac{\partial \{N^e\}^T}{\partial y}\mathrm{d}\Omega
\end{equation*}
 
\begin{equation*}
[A_{zx}^e]=[A_{xz}^e]^T=\iint_{\Omega^e}\frac{k_z}{\epsilon_r}\frac{\partial \{N^e\}}{\partial x}\{N^e\}^T\mathrm{d}\Omega
\end{equation*}

\begin{equation*}
[B_x^e]=[B_y^e]=[B_z^e]=\iint_{\Omega^e}\mu_r\{N^e\}\{N^e\}^T\mathrm{d}\Omega
\end{equation*} 
For simplicity we denote the matrices $[C_{xy}^e]\!=\![C_{yx}^e]^T, [D_x^e],~\text{and}~[D_y^e]$ having the matrix elements 
\begin{eqnarray}
\label{eq:17}
C_{xyij}^e&{}={}&\iint_{\Omega^e}\frac{\partial N_i^e}{\partial x}\frac{\partial N_j^e}
{\partial y}\mathrm{d}x\mathrm{d}y,  \qquad
D_{xij}^e = \iint_{\Omega^e}N_i^e\frac{\partial N_j^e}
{\partial x}\mathrm{d}x\mathrm{d}y   \nonumber \\
D_{yij}^e&{}={}&\iint_{\Omega^e}N_i^e\frac{\partial N_j^e}{\partial y}\mathrm{d}x\mathrm{d}y,\qquad i,j{}={}1,2,\dots 10. 
\end{eqnarray}
Explicit expressions for the matrix elements $C_{xyij}^e$ and $D_{xij}^e$ are given in Tables \ref{tab:3} and
\ref{tab:4}, respectively. In Table \ref{tab:3} we used notation $p_{ij}\!=\!b_ic_j$, with $i,j\!=\!1,2,3$. The matrix elements $D_{yij}^e$ are obtained by changing $b_i$ to $c_i~(i\!=\!1,2,3)$ in Table \ref{tab:4}.

\begin{table}[!ht]
\begin{center}
\caption{\label{tab:3}Matrix elements $C_{xyij}^e~(i,j=1,2,\dots 10)$ are tabulated expressions multiplied by $[81/(16\Delta^e)]$}
\begin{tabular}{|l||c|c|c|c|c|}
\hline
\hline
$i\backslash j$&\textit{1}&\textit{2}&\textit{3}&\textit{4}&\textit{5}\\
\hline
\hline
\textit{1}&\mbox{$\frac{17p_{11}}{405}$}&\mbox{$\frac{7p_{12}}{810}$}&\mbox{$\frac{7p_{13}}{810}$}&\mbox{ $\frac{18p_{12}-p_{13}}{270}$}&\mbox{$\frac{-p_{13}-9p_{12}}{270}$}\\
\hline
\textit{2}&\mbox{$\frac{7p_{21}}{810}$}&\mbox{$\frac{17p_{22}}{405}$}&\mbox{$\frac{7p_{23}}{810}$}&\mbox{ $\frac{-p_{23}-9p_{21}}{270}$}&\mbox{$\frac{18p_{21}-p_{23}}{270}$}\\
\hline 
\textit{3}&\mbox{$\frac{7p_{31}}{810}$}&\mbox{$\frac{7p_{32}}{810}$}&\mbox{$\frac{17p_{33}}{405}$}&\mbox{ $\frac{-p_{33}}{270}$}&\mbox{$\frac{-p_{33}}{270}$}\\
\hline
\textit{4}&\mbox{$\frac{18p_{21}-p_{31}}{270}$}&\mbox{$\frac{-p_{32}-9p_{12}}{270}$}&\mbox{$\frac{-p_{33}}{270}$}
&\mbox{$\frac{p_{11}+p_{22}+p_{33}}{12}$}&\mbox{$\frac{5p_{12}-p_{21}-2(p_{11}+p_{22})}{60}$}\\
\hline 
\textit{5}&\mbox{$\frac{-p_{31}-9p_{21}}{270}$}&\mbox{$\frac{18p_{12}-p_{32}}{270}$}&\mbox{$\frac{-p_{33}}{270}$} &\mbox{$\frac{5p_{21}-p_{12}-2(p_{11}+p_{22})}{60}$}&\mbox{$\frac{p_{11}+p_{22}+p_{33}}{12}$}\\
\hline
\textit{6}&\mbox{$\frac{-p_{11}}{270}$}&\mbox{$\frac{18p_{32}-p_{12}}{270}$}&\mbox{$\frac{-p_{13}-9p_{23}}{270}$}&\mbox{$\frac{-p_{13}-p_{31}}{60}$}&\mbox{$\frac{p_{13}+p_{31}}{12}$}\\
\hline 
\textit{7}&\mbox{$\frac{-p_{11}}{270}$}&\mbox{$\frac{-p_{12}-9p_{32}}{270}$}&\mbox{$\frac{18p_{23}-p_{13}}{270}$}&\mbox{$\frac{-p_{13}-p_{31}}{60}$}&\mbox{$\frac{-p_{13}-p_{31}}{60}$}\\
\hline
\textit{8}&\mbox{$\frac{-p_{21}-9p_{31}}{270}$}&\mbox{$\frac{-p_{22}}{270}$}&\mbox{$\frac{18p_{13}-p_{23}}{270}$}&\mbox{$\frac{-p_{23}-p_{32}}{60}$}&\mbox{$\frac{-p_{23}-p_{32}}{60}$}\\
\hline
\textit{9}&\mbox{$\frac{18p_{31}-p_{21}}{270}$}&\mbox{$\frac{-p_{22}}{270}$}&\mbox{$\frac{-p_{23}-9p_{13}}{270}$}&\mbox{$\frac{p_{23}+p_{32}}{12}$}&\mbox{$\frac{-p_{23}-p_{32}}{60}$}\\
\hline
\textit{10}&\mbox{0}&\mbox{0}&\mbox{0}&\mbox{$\frac{p_{13}+p_{31}}{10}$}&\mbox{$\frac{p_{23}+p_{32}}{10}$}\\
\hline
\hline
$i\backslash j$&\textit{6}&\textit{7}&\textit{8}&\textit{9}&\textit{10}\\
\hline
\hline
\textit{1}&\mbox{$\frac{-p_{11}}{270}$}&\mbox{$\frac{-p_{11}}{270}$}&\mbox{ $\frac{-p_{12}-9p_{13}}{270}$}&\mbox{$\frac{18p_{13}-p_{12}}{270}$}&\mbox{0}\\
\hline
\textit{2}&\mbox{$\frac{18p_{23}-p_{21}}{270}$}&\mbox{$\frac{-p_{21}-9p_{23}}{270}$}&\mbox{$\frac{-p_{22}}{270}$}&\mbox{$\frac{-p_{22}}{270}$}&\mbox{0}\\
\hline
\textit{3}&\mbox{$\frac{-p_{31}-9p_{32}}{270}$}&\mbox{$\frac{18p_{32}-p_{31}}{270}$}&\mbox{$\frac{18p_{31}-p_{32}}{270}$}&\mbox{$\frac{-p_{32}-9p_{31}}{270}$}&\mbox{0}\\
\hline
\textit{4}&\mbox{$\frac{-p_{13}-p_{31}}{60}$}&\mbox{$\frac{-p_{13}-p_{31}}{60}$}&\mbox{$\frac{-p_{23}-p_{32}}{60}$}&\mbox{$\frac{p_{23}+p_{32}}{12}$}&\mbox{$\frac{p_{13}+p_{31}}{10}$}\\
\hline
\textit{5}&\mbox{$\frac{p_{13}+p_{31}}{12}$}&\mbox{$\frac{-p_{13}-p_{31}}{60}$}&\mbox{$\frac{-p_{23}-p_{32}}{60}$}&\mbox{$\frac{-p_{23}-p_{32}}{60}$}&\mbox{$\frac{p_{23}+p_{32}}{10}$}\\
\hline
\textit{6}&\mbox{$\frac{p_{11}+p_{22}+p_{33}}{12}$}&\mbox{$\frac{5p_{23}-p_{32}-2(p_{22}+p_{33})}{60}$}&\mbox{$\frac{-p_{12}-p_{21}}{60}$}&\mbox{$\frac{-p_{12}-p_{21}}{60}$}&\mbox{$\frac{p_{12}+p_{21}}{10}$}\\
\hline
\textit{7}&\mbox{$\frac{5p_{32}-p_{23}-2(p_{22}+p_{33})}{60}$}&\mbox{ $\frac{p_{11}+p_{22}+p_{33}}{12}$}&\mbox{$\frac{p_{12}+p_{21}}{12}$}&\mbox{$\frac{-p_{12}-p_{21}}{60}$}
&\mbox{$\frac{p_{13}+p_{31}}{10}$}\\
\hline
\textit{8}&\mbox{$\frac{-p_{12}-p_{21}}{60}$}&\mbox{$\frac{p_{12}+p_{21}}{12}$}&\mbox{$\frac{p_{11}+p_{22}+p_{33}}{12}$}&\mbox{$\frac{5p_{31}-p_{13}-2(p_{11}+p_{33})}{60}$}&\mbox{$\frac{p_{23}+p_{32}}{10}$}\\
\hline
\textit{9}&\mbox{$\frac{-p_{12}-p_{21}}{60}$}&\mbox{$\frac{-p_{12}-p_{21}}{60}$} &\mbox{$\frac{5p_{13}-p_{31}-2(p_{11}+p_{33})}{60}$} &\mbox{$\frac{p_{11}+p_{22}+p_{33}}{12}$}&\mbox{$\frac{p_{12}+p_{21}}{10}$}\\
\hline
\textit{10}&\mbox{$\frac{p_{12}+p_{21}}{10}$}&\mbox{$\frac{p_{13}+p_{31}}{10}$}&\mbox{ $\frac{p_{23}+p_{32}}{10}$}&\mbox{$\frac{p_{12}+p_{21}}{10}$}&\mbox{$\frac{p_{11}+p_{22}+p_{33}}{5}$}\\
\hline
\hline
\end{tabular}
\end{center}
\end{table}

\begin{table}[!ht]
\begin{center}
\caption{\label{tab:4}Matrix elements $D_{xij}^e~(i,j=1,2,\dots 10)$ are tabulated expressions multiplied by $(27/140)$}
\begin{tabular}{|l||c|c|c|c|c|c|c|c|c|c|}
\hline
\hline
$i\backslash j$&\textit{1}&\textit{2}&\textit{3}&\textit{4}&\textit{5}&\textit{6}&\textit{7}&\textit{8}&\textit{9}&\textit{10}\\
\hline
\hline
\textit{1}&\mbox{$\frac{16b_1}{81}$}&\mbox{$\frac{19b_2}{324}$}&\mbox{$\frac{19b_3}{324}$}&\mbox{$\frac{22b_2-b_1}{72}$}&\mbox{$\frac{5b_1-8b_2}{72}$}&\mbox{$\frac{-5b_1}{72}$}&\mbox{$\frac{-5b_1}{72}$}
&\mbox{$\frac{5b_1-8b_3}{72}$}&\mbox{$\frac{22b_3-b_1}{72}$}&\mbox{$\frac{b_1}{12}$}\\
\hline
\textit{2}&\mbox{$\frac{19b_1}{324}$}&\mbox{$\frac{16b_2}{81}$}&\mbox{$\frac{19b_3}{324}$}&\mbox{$\frac{5b_2-8b_1}{72}$}&\mbox{$\frac{22b_1-b_2}{72}$}&\mbox{$\frac{22b_3-b_2}{72}$}&\mbox{$\frac{5b_2-8b_3}{72}$}
&\mbox{$\frac{-5b_2}{72}$}&\mbox{$\frac{-5b_2}{72}$}&\mbox{$\frac{b_2}{12}$}\\
\hline
\textit{3}&\mbox{$\frac{19b_1}{324}$}&\mbox{$\frac{19b_2}{324}$}&\mbox{$\frac{16b_3}{81}$}&\mbox{$\frac{-5b_3}{72}$}&\mbox{$\frac{-5b_3}{72}$}&\mbox{$\frac{5b_3-8b_2}{72}$}&\mbox{$\frac{22b_2-b_3}{72}$}&
\mbox{$\frac{22b_1-b_3}{72}$}&\mbox{$\frac{5b_3-8b_1}{72}$}&\mbox{$\frac{b_3}{12}$}\\
\hline
\textit{4}&\mbox{$\frac{23b_1}{72}$}&\mbox{$\frac{-13b_2}{72}$}&\mbox{$\frac{5b_3}{72}$}&\mbox{$-b_3$}&\mbox{ $\frac{b_2-3b_1}{8}$}&\mbox{$\frac{2b_1-b_3}{8}$}&\mbox{$\frac{b_1-b_3}{8}$}&\mbox{$\frac{b_2-2b_3}{8}$}&
\mbox{$\frac{b_3-b_2}{2}$}&\mbox{$\frac{2b_3-3b_1}{4}$}\\
\hline
\textit{5}&\mbox{$\frac{-13b_1}{72}$}&\mbox{$\frac{23b_2}{72}$}&\mbox{$\frac{5b_3}{72}$}&\mbox{$\frac{b_1-3b_2}{8}$}&\mbox{$-b_3$}&\mbox{$\frac{b_3-b_1}{2}$}&\mbox{$\frac{b_1-2b_3}{8}$}&\mbox{$\frac{b_2-b_3}{8}$}&\mbox{$\frac{2b_2-b_3}{8}$}&\mbox{$\frac{2b_3-3b_2}{4}$}\\
\hline
\textit{6}&\mbox{$\frac{5b_1}{72}$}&\mbox{$\frac{23b_2}{72}$}&\mbox{ $\frac{-13b_3}{72}$}&\mbox{$\frac{b_3-2b_1}{8}$}&\mbox{$\frac{b_1-b_3}{2}$}&\mbox{$-b_1$}&\mbox{ $\frac{b_3-3b_2}{8}$}&\mbox{$\frac{2b_2-b_1}{8}$}&\mbox{$\frac{b_2-b_1}{8}$}&\mbox{$\frac{2b_1-3b_2}{4}$}\\
\hline
\textit{7}&\mbox{$\frac{5b_1}{72}$}&\mbox{$\frac{-13b_2}{72}$}&\mbox{$\frac{23b_3}{72}$}&\mbox{$\frac{b_3-b_1}{8}$}&\mbox{$\frac{2b_3-b_1}{8}$}&\mbox{$\frac{b_2-3b_3}{8}$}&\mbox{$-b_1$}&\mbox{$\frac{b_1-b_2}{2}$}
&\mbox{$\frac{b_2-2b_1}{8}$}&\mbox{$\frac{2b_1-3b_3}{4}$}\\
\hline
\textit{8}&\mbox{$\frac{-13b_1}{72}$}&\mbox{$\frac{5b_2}{72}$}&\mbox{$\frac{23b_3}{72}$}&\mbox{$\frac{2b_3-b_2}{8}$}&\mbox{$\frac{b_3-b_2}{8}$}&\mbox{$\frac{b_1-2b_2}{8}$}&\mbox{$\frac{b_2-b_1}{2}$}&\mbox{$-b_2$}
&\mbox{$\frac{b_1-3b_3}{8}$}&\mbox{$\frac{2b_2-3b_3}{4}$}\\
\hline
\textit{9}&\mbox{$\frac{23b_1}{72}$}&\mbox{$\frac{5b_2}{72}$}&\mbox{ $\frac{-13b_3}{72}$}&\mbox{$\frac{b_2-b_3}{2}$}
&\mbox{$\frac{b_3-2b_2}{8}$}&\mbox{$\frac{b_1-b_2}{8}$}&\mbox{$\frac{2b_1-b_2}{8}$}&\mbox{$\frac{b_3-3b_1}{8}$}&\mbox{$-b_2$}&\mbox{$\frac{2b_2-3b_1}{4}$}\\
\hline
\textit{10}&\mbox{$\frac{-b_1}{12}$}&\mbox{$\frac{-b_2}{12}$}&\mbox{$\frac{-b_3}{12}$}&\mbox{$\frac{3b_1-2b_3}{4}$}&\mbox{$\frac{3b_2-2b_3}{4}$}&\mbox{$\frac{3b_2-2b_1}{4}$}&\mbox{$\frac{3b_3-2b_1}{4}$}&
\mbox{$\frac{3b_3-2b_2}{4}$}&\mbox{$\frac{3b_1-2b_2}{4}$}&\mbox{0}\\
\hline
\hline
\end{tabular}
\end{center}
\end{table}

For a triangle with coordinates of vertices $(1,1),(2,1),~\text{and}~(1,2)$, the area is $\Delta^e\!=\!0.5$ and the maximum absolute values on the columns of these elemental matrices vary in the intervals: Max$|B_{ij}^e
|\!=\!0.0057-0.1446$, Max$|A_{xij}^e|\!=\!0-4.05$, Max$|C_{xyij}^e|\!=\!0-2.025$, Max$|D_{xij}^e|\!=\!0-0.2411$. We can see that the elements of matrix $[B^e]$ are very small. Thus, they could be more accurate than those obtained by performing the integrations numerically. Note that relations presented here for the matrix elements are more simple than those presented in \cite{3,6}.

\section{Examples}

In the following we consider some waveguiding problems which have analytic solutions. The most simple is the eigenvalue problem of a hollow square waveguide \cite{7}. In Table \ref{tab:5} we give the results for the wavenumber $k_0(\rm{cm}^{-1})$ obtained by using the vectorial finite element method \cite{2} in terms of the magnetic field components $[H_x, H_y, H_z]$, when the propagation constant on the $z$-axis direction, $k_z$ equals zero. An uniform, one-directional mesh is considered with second-order and third-order triangular elements in the domain $x,y\in[-0.5\rm{cm}~0.5\rm{cm}]$. For second-order triangles, the number of elements is $18$ and the total number of nodes is $49$. For third-order triangles, the corresponding numbers are $18$ and $100$. Explicit relations of matrices for the second-order triangular elements were given in \cite{8,9}. As seen in Table \ref{tab:5}, the results obtained using third-order elements have better accuracy than those obtained with the same number of second-order elements.

\begin{table}[!h]
\begin{center}
\caption{\label{tab:5}Results for the hollow square waveguide}
\begin{tabular}{|c|c|c|}
\hline
\hline
\mbox{Analytic}&\mbox{Second Order}&\mbox{Third Order}\\
\mbox{Solution}&\mbox{Triangles}&\mbox{Triangles}\\
\hline
\hline
3.1416&3.1438&3.1416\\
\hline
4.4429&4.4523&4.4431\\
\hline
6.2832&6.3451&6.2852\\
\hline
\hline
\end{tabular}
\end{center}
\end{table}

\begin{figure}[h]
\begin{center}
\includegraphics[width=2.5in]{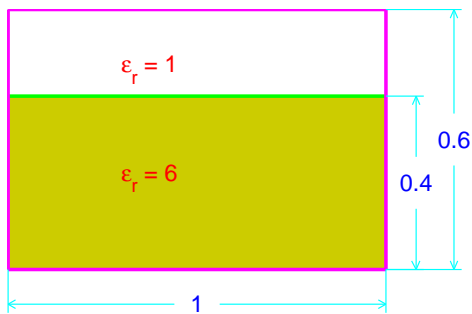}
\caption{Cross section of the inhomogeneous rectangular waveguide}
\label{fig:2}
\end{center}
\end{figure}

As another example we consider a rectangular waveguide partially filled with a dielectric \cite{7} of relative permittivity $\epsilon_r\!=\!6$ as illustrated in Fig.~\ref{fig:2}. Results for $k_0(\rm{cm}^{-1})$ are given comparatively in Table \ref{tab:6} at $k_z\!=\!0$ and $1\rm{cm}^{-1}$, the rectangular domain of $1\rm{cm}\times1\rm{cm}$ being uniformly divided in $24$ second-order elements with $63$ nodes or $24$ third-order elements with $130$ nodes.

\begin{table}[!ht]
\begin{center}
\caption{\label{tab:6}Results for the inhomogeneous rectangular waveguide}
\begin{tabular}{|c|c||c|c|}
\hline
\hline
&\mbox{Analytic}&\mbox{Second Order}&\mbox{Third Order}\\
\mbox{$k_z$}&\mbox{Solution}&\mbox{Triangles}&\mbox{Triangles}\\
\hline
\hline
0&1.7666&1.7681&1.7666\\
\hline
&2.3053&2.3076&2.3053\\
\hline
&2.6779&2.6875&2.6779\\
\hline
\hline
1&1.8310&1.8545&1.8531\\
\hline
&2.3460&2.3543&2.3521\\
\hline
&2.7125&2.7931&2.7042\\
\hline
\hline
\end{tabular}
\end{center}
\end{table}

\begin{table}[!ht]
\begin{center}
\caption{\label{tab:7}Results for the ferrite completely filled rectangular waveguide}
\begin{tabular}{|c|c|c|}
\hline
\hline
\mbox{Analytic}&\mbox{Second Order}&\mbox{Third Order}\\
\mbox{Solution}&\mbox{Triangles}&\mbox{Triangles}\\
\hline
\hline
0.6654&0.6659&0.6654\\
\hline
1.3307&1.3445&1.3307\\
\hline
1.9961&1.9458&1.9961\\
\hline
\hline
\end{tabular}
\end{center}
\end{table}

Finally we consider a rectangular waveguide with a $2\!:\!1$ width to height ratio completely filled with a ferrite material characterized by a relative permittivity $\epsilon_r\!=\!2$ and a relative permeability tensor $\boldsymbol{\mu_r}$ given by \cite{6}
\begin{equation}
\label{eq:18}
\boldsymbol{\mu_r}=\begin{bmatrix}3&0&j0.8\\0&1&0\\-j0.8&0&3
\end{bmatrix}.
\end{equation}
The analytical solution for the wave number $k_0$ of the $n$th mode is \cite{6}
\begin{equation}
\label{eq:19}
k_n^2=\frac{3.0}{16.72}\left[k_z^2+\left(n\frac{\pi}{2}\right)^2\right].
\end{equation}
Results are given comparatively in Table \ref{tab:7} for $k_0(\rm{cm}^{-1})$ of the first three modes, at $k_z\!=\!0$, when the rectangular domain of $2\rm{cm}\times1\rm{cm}$ is uniformly divided in $18$ second-order elements with $49$ nodes or in $18$ third-order elements with $100$ nodes. We can see that in the case of third-order elements the results agree with analytical solutions very well.

\end{document}